\documentstyle[11pt,paspconf]{article}

\begin{document}

\title{Limits on the cosmic infrared background from clustering in COBE/DIRBE
maps.}
\author{A. Kashlinsky$^{1,3}$, J. C. Mather$^2$, S. Odenwald$^3$}

\affil{$^1$ NORDITA, Blegdamsvej 17, Copenhagen DK-2100, Denmark}

\affil{$^2$ Code 685, NASA Goddard Space Flight Center, Greenbelt, MD 20771}

\affil{$^3$ Hughes STX, Code 685.3, NASA Goddard Space Flight Center}

\begin{abstract}
We discuss a new method of estimating the cosmic infrared background (CIB) from
the spatial properties of infrared maps and 
give the limits on the CIB from applying it to the
COBE/DIRBE maps. The strongest limits are obtained at mid- to far-IR
where foregrounds are bright, but smooth. If the CIB comes from matter
clustered like galaxies, the smoothness of the maps implies CIB levels
less than $\sim$(10-15) nW/m$^2$/sr over this wavelength range.
\end{abstract}


\section{Introduction}

The cosmic infrared background (CIB) contains information about the conditions
in the post-recombination Universe and links the microwave background,
which probes the last scattering surface, and the optical part of the
cosmic spectrum, which probes the conditions
in the Universe today at $z\sim 0$. The predicted spectral properties and 
the amplitude of the CIB depend on the
various
cosmological assumptions used, such as the cosmological density parameter,
the history of
star formation, and the power spectrum of the primordial density field among
others. A typical prediction over the range  of wavelengths probed by
the COBE/DIRBE, 1.25 - 240 $\mu$m, is $\nu I_\nu \sim 10$ nW/m$^2$/sr.  
It is difficult to measure such levels
directly because the foreground emissions from stars and interstellar
and interplanetary dust are bright.

An alternative to direct photometric measurements of the uniform (DC)
component
of the CIB is to study its spatial structure. In
such a method, the groundwork for
which has been laid in Kashlinsky, Mather, Odenwald and Hauser (1996, KMOH), 
one compares the predicted fluctuations of the CIB
with the measured angular correlation function of the maps. Here  we discuss the
results from applying this method to the all-sky DIRBE maps and show that
it imposes interesting limits on the CIB from clustered matter,
particularly in the mid- to far-IR (Kashlinsky, Mather and Odenwald 1996; KMO).

\section{Theoretical expectations}

In general, the
intrinsic correlation function of the diffuse
background, $C(\theta) \equiv \langle \nu \delta I_\nu({\bf x}) \cdot
\nu \delta I_\nu ({\bf x}+\theta)\rangle$, where $\delta I \equiv I-
\langle I \rangle$ is the map of spatial fluctuations, depends on the 3-dimensional correlation
function $\xi(r)$ of the emitters, the history of their emission and cosmological
parameters. The correlation function of the diffuse background, $C(\theta)$
produced by a population 
of emitters (e.g. galaxies) clustered with a 2-dimensional correlation function $\xi(r)$ is given 
in the small angle limit ($\theta < 1$) by:
\begin{equation}
C(\theta) = \int_0^\infty A_\theta(z) (\frac{d\lambda I_\lambda}{dz})^2 [\Psi^2(z) (1+z)^2
\sqrt{1+\Omega z}] dz
\end{equation}
where it was assumed that the cosmological constant is zero, $\Psi(z)$ accounts
for the evolution of the clustering pattern, 
$A_\theta (z)$=$2R_H^{-1} \int_0^\infty
\xi(\sqrt{v^2 + \frac{x^2(z)}{(1+z)^2}}) dv$, $R_H$$=$$cH_0^{-1}$ and $x(z)$ is the comoving 
distance.  Convolving (1)
with the beam leads to the zero-lag signal for the DIRBE sky:
\begin{equation}
C_\vartheta(0) = \int_0^\infty A_\vartheta(z) (\frac{d\lambda I_\lambda}{dz})^2 [\Psi^2 (1+z)^2
\sqrt{1+\Omega z}] dz
\end{equation}
\begin{equation}
A_\vartheta(z)=\frac{1}{2\pi R_H} \int_0^\infty P_{3,0}(k) k W(\frac{kx(z)\vartheta}{1+z}) dk
\end{equation}
For the top-hat beam of DIRBE $W(x)=[2J_1(x)/x]^2$ and $\vartheta$=0.46\deg. In the
above equation $P_{3,0}(k)$ is the spectrum of galaxy clustering at the present epoch. If $\xi$  is 
known the measurement of $C(\theta)$ can give 
information on the diffuse background (Gunn 1965). Such method was
applied in the V (Shectman 1973,1974) and UV bands (Martin and Bowyer 1989). 
In the rest of the paper
we omit the subscript $\vartheta$ in (2) with $C(0)$ referring to the DIRBE 
convolved zero-lag correlation signal.

The functional form of $A_\vartheta(z)$, for the \underline{data} 
on the present
day power spectrum $P_{3,0}(k)$ taken from galaxy catalogs, 
is discussed in KMO and it is shown that
it reaches a minimum at $z$$\geq$1. 
The minimal value of $A_\vartheta(z)$ for the scales 
probed by the DIRBE beam is $\sim 10^{-2}$ and is practically independent 
of $\Omega$. Thus we can rewrite (2) as an inequality in order to derive 
upper limit on a measure of the CIB flux from
any upper limit on $C(0)$ derived from the DIRBE data:
$(\lambda I_\lambda)_{z,rms}$$\leq$$B \sqrt{C(0)}$
where $B$$\equiv$$1/\sqrt{{\rm min}\{A_\vartheta(z)\}}$=(11-13)
over the entire range of parameters and the
measure of the CIB flux used in (4) is defined as 
$[(\lambda I_\lambda)_{z,rms}]^2$$\equiv$$\int 
(\frac{d\lambda I_\lambda}{dz})^2 [\Psi^2(z)(1+z)^2\sqrt{1+\Omega z}] dz$. 
The latter is $\simeq$$\int (\frac{d\lambda I_\lambda}{dz})^2 dz$ 
since the term in the brackets has little variation with $z$ for two
extremes of clustering evolution when 
it is stable in either proper or comoving coordinates.

Thus for scales probed by DIRBE,
the CIB  produced by objects clustered like
galaxies should have significant fluctuations, $\geq 10\%$ of the total
flux, on the angular scale subtended by the DIRBE beam. The upper
limit on $C(0)$ would imply an upper limit on the total CIB from clustered
matter.

\section{Data analysis}

We analyzed maps for all 10 DIRBE bands derived from the entire
41 week DIRBE data set available from the NSSDC. A parametrized model
developed by the DIRBE team (Reach et al. 1996) was used to remove the
time-varying zodiacal component from each weekly map (KMOH).
The maps were pixelized using the quadrilateralized spherical cube projection
of the sky maps.
After the maps were constructed, the sky was divided into 384 patches of
32$\times$32 pixels $\simeq 10\deg\times 10\deg$  each. 

Each field was cleaned
of bright sources by the program developed by the DIRBE team and discussed
and used in KMOH, KMO.
The flux distribution in each patch was modelled with a smooth 4-th order 
polynomial component. Pixels with fluxes $> N_{\rm cut}$ standard 
deviations above the fitted model 
were removed along with the surrounding 8 pixels. Three values of 
$N_{\rm cut}$ were used: 
$N_{\rm cut}=7,5,3.5$. Since any large-scale gradients in the 
emission are clearly due to the local foregrounds,
a 3rd order polynomial was removed from each patch after desourcing. 
As discussed in Paper I we verified
that there is a good correlation between  removed bright objects at 
$N_{\rm cut}$ and the stars from the
SAO catalog; at lower $N_{\rm cut}$ the removed peaks would be 
too dim to enter the catalog.

Three types of foregrounds provide dominant contributions at various bands: 
in the near-IR bands (1-4) the
main foreground contribution to the zero-lag signal comes from the Galactic 
stars, in the mid-IR (bands
5-7) from the zodiacal light emission, and in bands 8-10 the main 
contribution is from the cirrus clouds in
the Galaxy. The structure of the near-IR maps and the distribution of 
fluctuations within the various fields
have been discussed at length in KMO; the histograms for these bands are 
shown there in Fig.5. They are
highly asymmetric with respect to $\delta I_\lambda$ and are consistent 
with those from simulated stellar 
catalogs. The distribution at mid- to far-IR bands
is more symmetric and comes from
a more extended emission. Furthermore, because of the extended character 
of the mid-to far-IR  foreground emission the decrease of the width of 
the histogram (another measure of $C(0)$) with $N_{\rm cut}$
is significantly less prominent than for near-IR where stars provide 
the dominant foreground. 

Removal of the large-scale gradients does not lead to any noticeable 
difference in the histograms at the
near-IR bands (1-4). On the other hand, at longer wavelengths the bulk 
of the foreground comes from 
extended emission and removing large-scale gradients, which are clearly 
of local origin, changes the structure significantly.

Histograms of the distribution of $C(0)$ on the sky show that 
in the near-IR where Galactic stars dominate the foreground fluctuations 
there is a significant decrease in the minimal value of $C(0)$  with $N_{\rm cut}$. The reason for this is 
that the structure of the foreground is dominated by fluctuations in point sources (stars) affecting the
small scale structure of the correlation function. In the near-IR the histograms become increasingly
asymmetric with decreasing $N_{\rm cut}$ and the bins with low $C(0)$ patches at higher clipping 
thresholds now have essentially the same $C(0)$. On the other hand, at mid- to far-IR there is very
little change in the distribution of the zero-lag signal with decreasing $N_{\rm cut}$; this is due to the
fact that the foreground emission is extended with little small-scale structure. The latter in turn leads to
much smaller upper limits on the value of $C(0)$ at mid- to far-IR. At bands 9,10 the instrumental noise
is substantially larger and the correlations analysis of the DIRBE maps leads to significantly less 
interesting limits at these wavelengths ($\lambda=150, 240\mu$m). 

We calculated the minimum of $C(0)$ from this analysis in all 10 bands.
The estimated upper limits 
come from the quietest patch in each band for the entire sky.
The histogram distribution of $C(0)$ shows that
the minimum is estimated from around 30-80 patches in each of the bands
making it a reliable estimate of the true upper limit. Since
the distribution of $C(0)$ is highly anisotropic on the sky, the signal must
come from the foreground emission and it is thus further unlikely that
any of the truly cosmological contributions have been removed in the
process. In the near IR these are located around
Galactic poles; in the mid-IR they are near Ecliptic poles and in the
far-IR they shift back to the Galactic poles. Our data processing does not
remove real fluctuations unless they appear to come from point sources,
or to have very large scale gradients.  The instrument digitization
noise, estimated to be below 1.0 nW/m$^2$/sr in all bands after averaging,
is not a problem at the level probed here.

\section{Conclusions}

We measured the smoothness of the infrared sky using the COBE DIRBE
maps, and obtain interesting limits on the production of the diffuse
cosmic infrared background (CIB) light by matter clustered like galaxies.
The predicted fluctuations of the CIB with the DIRBE beam size of 0.7$\deg$
are of the order of 10\%, and the maps are smooth at the level of
$\nu \delta I_\nu \sim $ a few nW/m$^2$/sr rms from 2.2 to 100 $\mu$m.
The lowest numbers are achieved at mid- to far-IR where the foreground
is bright but smooth; they are $\sqrt{C(0)}$$\leq$$(1-1.5)$ nW/m$^2$/sr at
$\lambda$=10-100 $\mu$m.  If the
CIB comes from clustered matter evolving according to typical
scenarios, then the smoothness of the maps implies CIB levels less than
$\sim$(10-15) nW/m$^2$/sr over this wavelength range.

\acknowledgments This work was supported in part by the NASA Long Term
Astrophysics grant, and we thank the many members of the COBE DIRBE team
whose work made this project possible. NASA Goddard Space Flight
Center is responsible for the design,
development and operation of COBE. GSFC is also responsible for the development
of analysis software and production of the mission data sets. The COBE program
is supported by the Astrophysics division of NASA's Office of Space Science 
and Applications.

\begin{question}{A. Franceschini}
Since the long wavelength detections ($\lambda = 60,100$ micron) are 
dominated by late type galaxies which are quite less clustered than optical
or near-IR galaxies, the correct limits then are probably somewhat higher
than appearing in the figure.
\end{question}
\begin{answer}{}
It is true that radio galaxies are clustered more weakly on scales where
density field is non-linear ($<5h^{-1}$Mpc). However, 
 at the relevant range of redshifts the DIRBE beam of
roughly 1 degree
subtends scales which are today in linear 
or quasilinear regime of clustering
and where one would expect that all galaxies probe the same density field.
Our correlation amplitude estimates use as an input 
the data on the correlation
function of blue (APM) and red (POSS) galaxies. Most contribution to the
CIB fluctuations at 
1 degree comes from scales $\simeq (10-30)h^{-1}$Mpc where
(presumably) radio or IR galaxies would be distributed similarly. 
But the effect you
are referring to may indeed be significant in smaller beam surveys.
\end{answer}

\begin{question}{T. Matsumoto}
I suppose you can estimate the fluctuation of stars using lg$N$/lg$S$
plot, and can obtain lower fluctuation limit for CIB.
\end{question}
\begin{answer}{}
Yes, the fluctuation in stars can be estimated from the lg$N$/lg$S$ plots
and we find the estimates to be in agreement with the direct computation.
But we did not succeed in obtaining robust model-independent limits on the CIB
in this or similar ways. 
\end{answer}


\begin{references}

\reference Gunn, J. 1965, Ph.D. Thesis, Caltech. (unpublished)

\reference Kashlinsky, A., Mather, J., Odenwald, S. and Hauser, M. 1996, Ap.J.,
{\bf 470},681. (KMOH)

\reference Kashlinsky, A., Mather, J., Odenwald, S. 1996, Ap.J.,
{\bf 473},L9. (KMO)

\reference Martin, C. and Bowyer, S. 1989, Ap.J, {\bf 338}, 677.

\reference Reach, W. et al, 1996, in ``Unveiling
the Cosmic Infrared Background," 1996, AIP Conf. Proc., {\bf 348}, (AIP:
New York), 37-46.

\reference Shectman, S. 1973, Ap.J., {\bf 179}, 681.

\reference Shectman, S. 1974, Ap.J., {\bf 188}, 233.

\end{references}
\end{document}